\documentclass[preprint, 10pt,authoryear]{elsarticle}
\usepackage{graphicx,graphics}
\usepackage{amsmath, amssymb, amsthm}
\usepackage{natbib}
\usepackage{color}
\usepackage[colorlinks=true,linkcolor=black, citecolor=blue, urlcolor=blue]{hyperref}
\usepackage{a4wide}
\usepackage[font=scriptsize,labelfont=bf]{caption}
\usepackage[draft]{changes}
\usepackage{lipsum}
\usepackage{subfigure}

\usepackage{tikz}
\usetikzlibrary{topaths}

\makeatletter
\def\ps@pprintTitle{%
 \let\@oddhead\@empty
 \let\@evenhead\@empty
 \def\@oddfoot{}%
 \let\@evenfoot\@oddfoot}
\makeatother

\newcommand{\mat}[1]{\mathbf{#1}}

\journal{Economics Letters}

\begin{document}

\begin{frontmatter}

\title{Extracting Geography from Trade Data}

\author[a]{Yuke Li}
\ead{yuke.li@yale.edu}
\author[b]{Tianhao Wu}
\ead{tianhao.wu@yale.edu}
\author[c]{Nicholas Marshall}
\ead{nicholas.marshall@yale.edu}
\author[d]{Stefan Steinerberger\corref{cor1}}
\ead{stefan.steinerberger@yale.edu}

\cortext[cor1]{Please address correspondence to Author 4.}
\address[a]{Department of Political Science, Yale University, New Haven, CT, USA}
\address[b]{Department of Statistics, Yale University, New Haven, CT, USA}
\address[c]{Department of Mathematics, Program in Applied Mathematics, Yale University, New Haven, CT, USA}
\address[d]{Department of Mathematics, Yale University, New Haven, CT, USA}

\date{\today}


\begin{abstract} 
Understanding international trade is a fundamental problem in economics -- one standard approach is via what is commonly called the ``gravity equation'', which
predicts the total amount of trade $F_ij$ between two countries $i$ and $j$ as
$$ F_{ij} = G \frac{M_i M_j}{D_{ij}},$$
where $G$ is a constant, $M_i, M_j$ denote the ``economic mass'' (often simply the gross domestic product) and $D_{ij}$ the ``distance'' between countries $i$ and $j$,
where ``distance'' is a complex notion that includes geographical, historical, linguistic and sociological components. We take the \textit{inverse} route and ask ourselves to
which extent it is possible to reconstruct meaningful information about countries simply from knowing the bilateral trade volumes $F_{ij}$: indeed, we show that a remarkable amount of geopolitical information can be extracted. The main tool is a spectral decomposition of the Graph Laplacian as a tool to perform nonlinear dimensionality
reduction. This may have further applications in economic analysis and provides a data-based approach to ``trade distance''. 

\end{abstract}


\end{frontmatter}
%

\section{Introduction} 

\subsection{The gravity equation.} 

Understanding the geometry and structural properties of world trade is a problem of obvious significance, great appeal and a long history. It has become challenging to summarize the existing literature outside of the framework of a survey article, a starting point is given by the (non-exhaustive) list~\citep{bhattacharya2008international, fagiolo2008topological,  fronczak2012statistical, garlaschelli2005structure, he2010structure,karpiarz2014international, li2003local,serrano2008rich, tzekina2008evolution} and references therein. One of the dominant paradigms within economics was first formalized in \citet{tinbergen1962shaping}. Given a set of $n$ countries, the ``gravity equation'' predicts the bilateral trade flow $F_{ij}$ between
countries $i$ and $j$ as
$$ F_{ij} = G \frac{M_i M_j}{D_{ij}},$$
where $G$ is a constant, $M_i, M_j$ denote the ``economic mass'' (often simply the gross domestic product) and $D_{ij}$ the ``distance'' between countries $i$ and $j$. ``Distance'' is a complex notion for which various models have been proposed -- these include factors such as common colonizers~\citep{frankel2002estimate}, cultural proximity~\citep{felbermayr2010cultural}, linguistic ties~\citep{melitz2014native}, mutual trust~\citep{butler2009right}, past conflicts~\citep{keshk2004trade}, shared borders~\citep{mccallum1995national}, shared currency~\citep{frankel2002estimate} and others. 

\subsection{Our approach.} 
Motivated by the complex notions of ``distance'' in the literature, we were interested in the question of whether it's possible to understand what type of underlying factors dominate multinational
trade \textit{using trade volume alone}: given merely the amount of trade $F_{ij}$ between different countries, how much information can be extracted about the ``distances'' 
between the countries? We believe that any such minimal approach, if successful, is most suited in furthering a precise understanding of the underlying processes.
Our approach is as follows: we formulate trade structure as complete, weighed graph. This graph is fairly complicated and not well-suited for a direct analysis -- however, we can ask whether it is possible to embed the graph in $\mathbb{R}^2$ without introducing a large distortion of distances. We will do this using \textit{diffusion maps} (see  \cite{coifman}): by performing
a spectral analysis of a suitable diffusion operator (the Graph Laplacian), we can use its ground state and first excited state as a nonlinear coordinate system.
Since the technique can also be applied to sub-graphs, this provides a map $$\phi: \left\{\mbox{collection of countries}\right\} \rightarrow \mathbb{R}^2$$ using only the total amount of trade done between any two countries in the collection within a given year. We show  that the map faithfully represents a complex notion of ``distance'' comprised of geographical
factors (different continents being easily identifiable), shared history, common language, colonization history and others.

\begin{figure}[h!]
\centering
\includegraphics[scale=0.52]{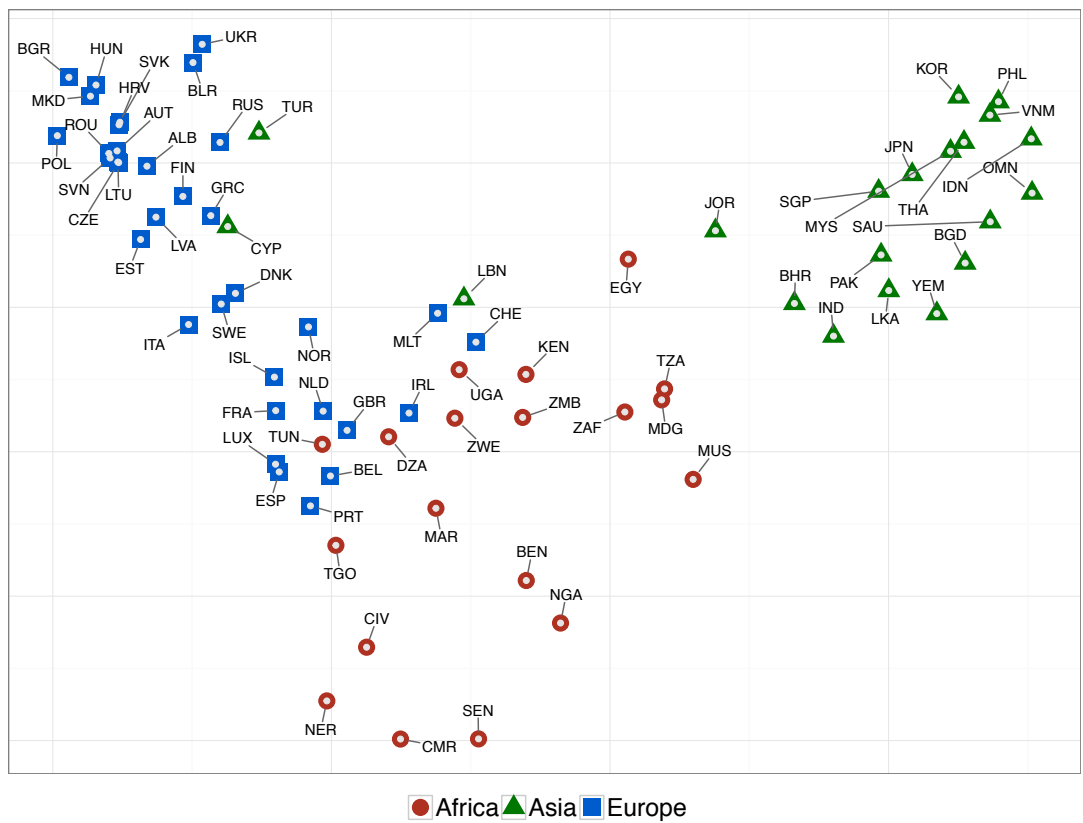}
\vspace{-4pt}
\caption{Applying the map to 73 countries using the COW world trade data set from the year of 2009.}
\label{2009aea}
\end{figure}

More precisely, the contribution of this paper is to construct an embedding into the Euclidean plane $\mathbb{R}^2$ that
\begin{enumerate}
\item automatically separates countries into different continents and meaningful subregions,
\item provides an accurate insight into what ``trade distance'' might be \textit{since the computation does not use any information about cultural, historical, linguistic or sociological matters},
\item shows that trade $(F_{ij})_{i,j = 1}^{n}$ has few underlying factors that dominate the global dynamics 
\item and, finally, it demonstrates that tools from nonlinear dimensionality reduction may be highly useful in the analysis of international trade.
\end{enumerate}

The striking difference between our approach and existing studies is as follows: first, to the best of our knowledge, our work is the first that applies nonlinear dimensionality reduction techniques in the studies of trade; secondly, in doing so, we obtain detailed and informative empirical regularities about selected countries' distances in trade \textit{without} any external information
regarding culture, geography, language or sociology: \textit{all information is solely derived from the trading behavior}.

\section{Mathematical Analysis}
\subsection{Setup}
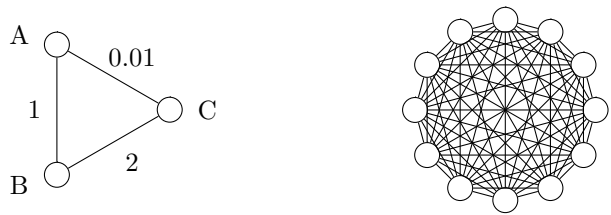
\begin{figure}[h!]
\minipage{0.5\textwidth}
\begin{tikzpicture}
\node at (-4,0) {};
 \def \radius {1cm}
 \def \margin {8}
 \def \n {3}
 \foreach \s in {1,...,\n}
  \node[draw, circle] (\s) at ({360/\n * (\s - 1)}:\radius) {};
 \foreach \s in {1,...,\n}
  \foreach \t in {\s,...,\n}
   \draw (\t) -- (\s);
\node at (0.5,0.7) {0.01};
\node at (-0.8,0) {1};
\node at (0.5,-0.7) {2};
\node at (-1,1) {A};
\node at (-1,-1) {B};
\node at (1.5,0) {C};
\end{tikzpicture}

\endminipage\hfill
\minipage{0.50\textwidth}
\begin{tikzpicture}
\node at (-2,0) {};
 \def \radius {1.2cm}
 \def \margin {8}
 \def \n {12}
 \foreach \s in {1,...,\n}
  \node[draw, circle] (\s) at ({360/\n * (\s - 1)}:\radius) {};
 \foreach \s in {1,...,\n}
  \foreach \t in {\s,...,\n}
   \draw (\t) -- (\s);
\end{tikzpicture}
\endminipage\hfill
\caption{A hypothetical scenario of 3 countries and their mutual trade flow (left). Generally, the problems are more complicated (right): this is a collection of 12 countries and the 132 directional trade flows.}
\label{ex}
\end{figure}

We proceed by interpreting the problem as one in dimensionality reduction: we map $n$ given countries to a complete, weighted graph
$$\phi_1:\left\{\mbox{collection of countries}\right\} \rightarrow \mbox{complete, weighted graph on}~n~\mbox{vertices}$$ 
that encodes the given information. The crucial part is the construction of a map 
$$\phi_2:\mbox{complete, weighted graph on}~n~\mbox{vertices} \rightarrow \mathbb{R}^2$$
that preserves as much information as possible. The composition $$\phi_2 \circ \phi_1:\left\{\mbox{collection of countries}\right\}  \rightarrow \mathbb{R}^2$$ is then the desired object.
We observe that the construction of $\phi_2$ is highly nontrivial since it is generally impossible to give such maps
without introducing enormous distortions in the underlying distances: let us consider (see Figure~\ref{ex}) an example of 3 countries $A,B,C$ such that the total trade flow between $A -B$, $B - C$ and
$C - A$ is $1, 2$ and $0.01$, respectively. 
Any map $\phi:\left\{A,B,C\right\} \rightarrow \mathbb{R}^2$ is going to map the three countries to three points in $\mathbb{R}^2$ and we would consider that map to be a faithful
representation if the pairwise distances are somewhat comparable to the inverse of the trade flow (such that two countries with large common trade flow are mapped next to each
other). This means we would like $A$ and $B$ as well as $B$ and $C$ to be mapped in close vicinity while simultaneously ensuring a large distance between $A,C$. \\

This is
obviously impossible; in larger examples, there are disproportionately more ways to create contradicting data. We see that the problem is generally not solvable --  \textit{unless
there is a strong underlying two-dimensional structure} encoded in the graph. This seems to indeed be the case for trade data: we believe that economists have long
suspected this and consider the search for an approximate notion of ``distance'' to be motivated by that suspicion.

\subsection{Setting up the maps} We use the diffusion embedding as introduced by \cite{coifman}. A rough description of the underlying idea is as follows: let us consider diffusion on a domain $\Omega \subset \mathbb{R}^2$ 
as modeled by the heat equation 
$$ \frac{\partial}{\partial t} u = \Delta u,$$
where $\Delta$ is the usual Laplacian and we impose Neumann conditions on the boundary. Suppose we are interested in the ground state, i.e. the profile which
decays the slowest under the heat equation: the ansatz $u(x,t) = e^{-\lambda t} \phi(x)$ gives
$$ -\Delta u = \lambda u$$
and requires us to find the smallest eigenfunction of $-\Delta$. 

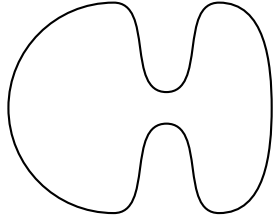
\begin{figure}[h!]

\begin{center}
\begin{tikzpicture}[scale= 0.7]
\draw [thick] (0,0) to [out=270,in=180] (2,-2) to [out=0,in=180] (3,-0.3) to [out=0,in=180] (4,-2) to [out=0, in=270] (5,0) 
to [out=90, in = 0] (4,2) to [out=180, in = 0] (4,2) to [out=180, in = 0] (3,0.3) to [out=180, in = 0] (2,2) to [out=180, in = 90] (0,0)   ;
\end{tikzpicture}
\end{center}
\caption{A domain with a bottleneck.}
\label{bottle}
\end{figure}

However, on domains as pictured in Figure \ref{bottle}, physical intuition tells us immediately what this lowest eigenfunction has to look like: it will be essentially constant on both sides of the
domain and undergo a rapid transition in the bottleneck. Physically speaking, the best way to separate high and low temperatures in a domain like $\Omega$ is to distribute them in such a way
that the actual surface where they meet is as small as possible. However, this just means that
$$ \phi:\Omega \rightarrow \mathbb{R}$$
is a good classifier to understand whether a point is on the left or on the right side of the bottleneck (as they will have opposite signs). We will now emulate this intuition on a complete, weighted
graph. Given a set of $n$ countries, let each country be represented by a unique element in $\{1,...,n\}$. We encode trade in a non-symmetric $\mat{W} = [w_{ij}]_{n \times n}$, where $w_{ij} \in \mathbb{R}^{+} \cup \left\{0\right\}$ is the total export from country $i$ to $j$; trivially, $w_{ii} = 0$. We only work with the total trade volume between countries $i$ and $j$ and use the \textit{affinity matrix} $\mat{A} = [a_{ij}]_{n\times n}$ defined by
$$ \mat{A} = \mat{W} + \mat{W}^T.$$ 
$\mat{A}$ encodes the weight of the edge between the vertices $i$ and $j$ in the complete, weighted graph.
The construction of $\phi_2$ is proceeds as follows:
\begin{enumerate}
\item We define a diagonal matrix $\mat{D} = [d_{ij}]_{n \times n}$, where $d_{ij}=0$ if $i\neq j$ and $$d_{ii}=\sum_{j=1}^{n} a_{ij}$$
\item Finally, we construct the normalized Laplacian matrix to be 
$$\mat{N} = \mat{I} - \mat{D}^{-1/2}\mat{A}\mat{D}^{-1/2},$$
where $\mat{I}$ is the identity matrix. We then compute the two eigenvectors $\textbf{v}_1, \textbf{v}_2$ associated with the smallest two nontrivial eigenvalues.
\item The map $\phi_2$ is given by using the inner product with these two eigenvectors 
$$ \phi_2(i) := ( (\textbf{v}_1)_i, (\textbf{v}_2)_i ) \in \mathbb{R}^2,$$
where $\textbf{v}_i$ denotes the $i-$th entry of a vector.
\end{enumerate}

\section{Results}
\subsection{Africa, Asia \& Europe.}
We first give an example of applying this method to all countries in Africa, Asia and Europe for which all trade flows are known and contained in the COW trade dataset~\citep{BarCorrelates, BarTrade} in the year 2009 (see Figure \ref{2009aea} above) -- this leads to a total of 73 countries being included and excludes two major economic contributors, China and Germany, for which the data is not fully available; there are several different ways to include countries with incomplete data (using data completion and extrapolation methods) but this is outside of the scope of this paper. 
Given the highly non-local nature of trade, the map recovers an astonishing number of meaningful features connected to various existing notions of ``distance''. Remarkably, there is an automatic separation of three 
continents -- there are a couple of outliers that make it an imperfect classification, however, these outliers are actually highly meaningful and as will be discussed, have 
been correctly placed.

\begin{figure}[h]
\begin{minipage}{0.44\textwidth}
\includegraphics[width=1.25\textwidth]{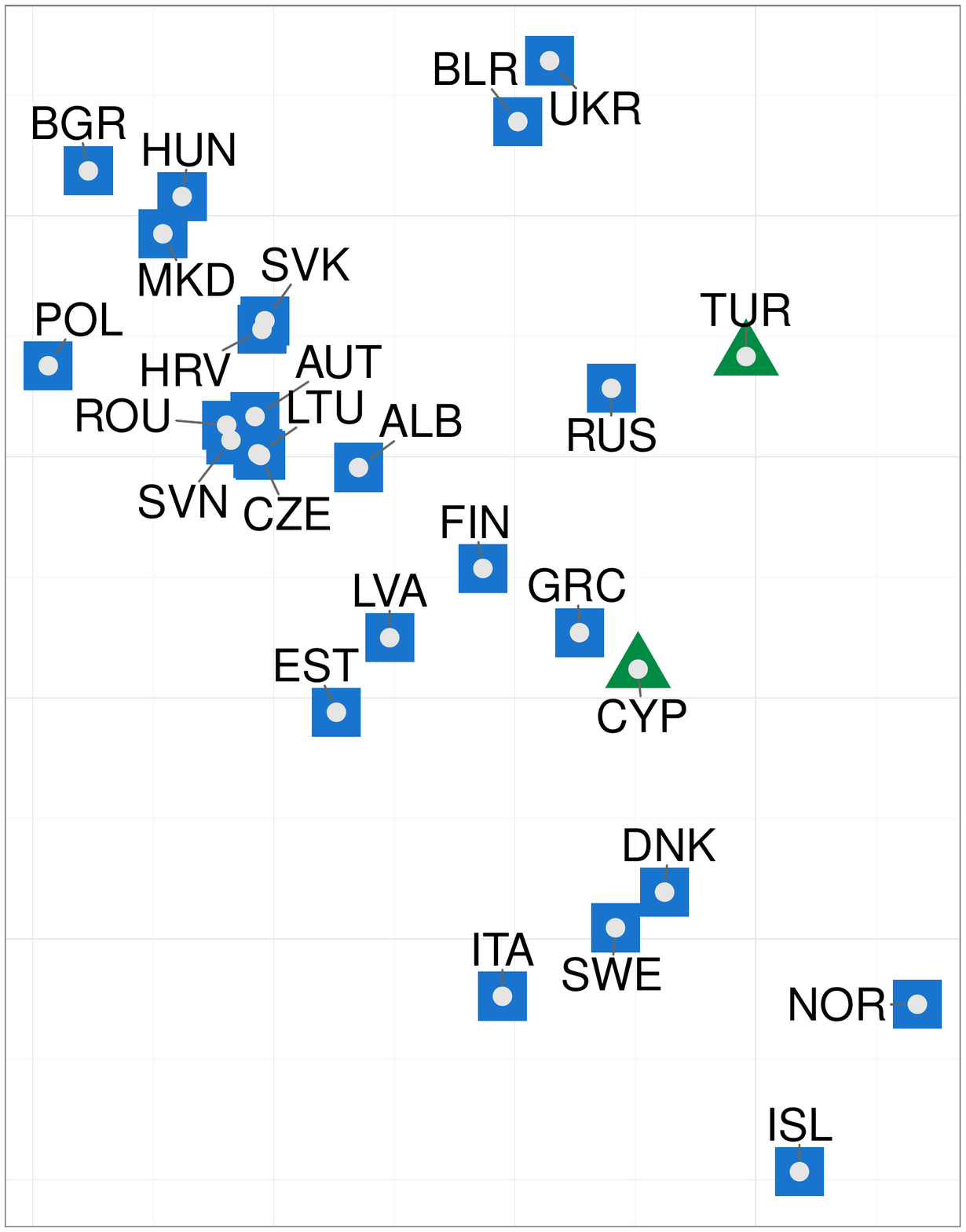}
\caption{Turkey and Cyprus' European Affinity}
\label{turkey}
\end{minipage}
\hfill
\begin{minipage}{0.53\textwidth}
\includegraphics[width=1.0\textwidth]{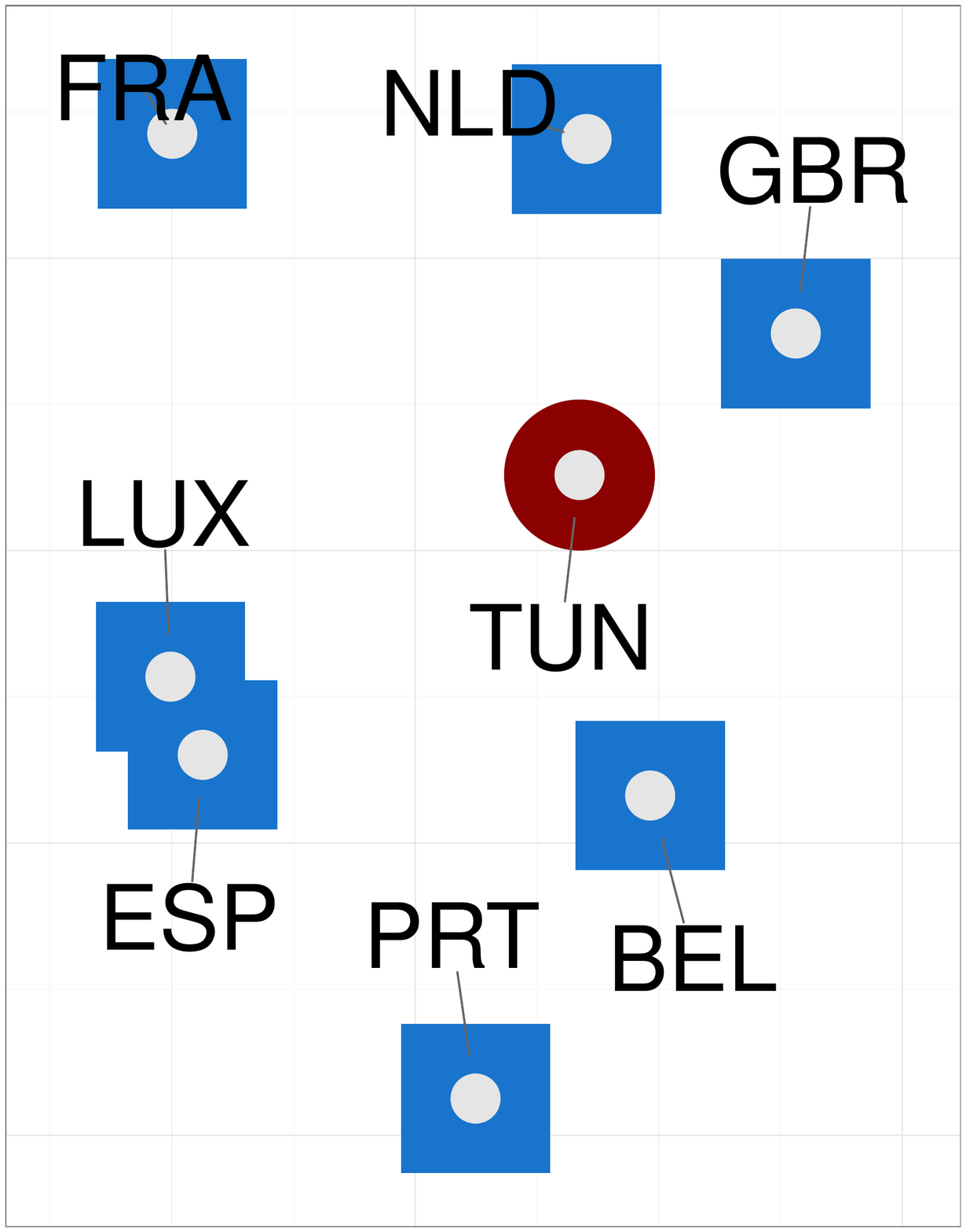}
\caption{Tunisia and its Neighbors}
\label{tunisia}
\end{minipage}
\end{figure}

Figure~\ref{turkey} and Figure~\ref{tunisia} show particularly interesting regions in Figure~\ref{2009aea}.
Cyprus is mapped as being very close to both Greece and Turkey, which is unsurprising from a historical
perspective. Austria is in a cluster surrounded by the Czech republic, Croatia, Hungary, Romania, Slovakia and Slovenia (all were at least partially elements of the Austro-Hungarian empire). Denmark, Norway and Sweden are in close proximity, and Finland is mapped next to the Baltic 
states Estonia, Latvia, Lithuania as well as Russia. Tunisia is placed in the middle of European countries: among its closest neighbors are Spain and Portugal
(geographic proximity), Belgium (linguistic proximity) and France (both historical and linguistic proximity). The same cluster also contains the Netherlands and Luxembourg as neighbors
of France (geographic, linguistic and historical proximity) and maps Great Britain as being close to France.

\subsection{Americas, Asia and Europe.} 

\begin{figure}[h!]
\includegraphics[width=0.95\textwidth]{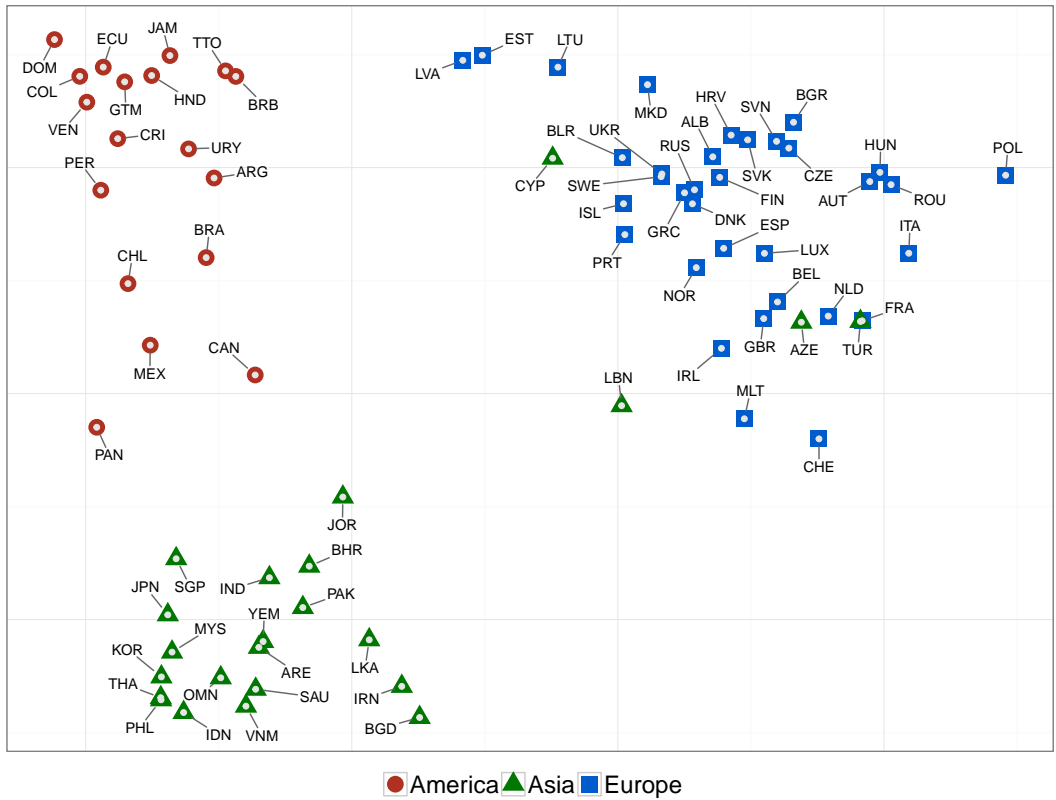}
\centering
\captionof{figure}{Americas, Asia and Europe}
\label{ex7}
\end{figure}

Figure~\ref{ex7} shows the map applied to America, Asia and Europe. We still observe a clear classification of continents but some 
subtle differences in the representation of Europe: Cyprus and Turkey are still classified as European countries. So are Azerbaijan and Lebanon which were already close
to Europe in Figure~\ref{2009aea} and have now moved to even closer to Europe: in the context of trading among America-Asia-European
countries, Azerbaijan, Cyprus, Lebanon and Turkey actively trade as if they were European countries. Observe that Switzerland (CHE) appears as a European
outlier and Panama is an American outlier. We emphasize that the method really constructs a map
$$ \phi: \left\{\mbox{collection of countries} \right\} \rightarrow \mathbb{R}^2$$
that tries to most appropriately describe the relationship among these countries and these countries alone --  ad hoc factors such
as the number of common high-volume trading partners outside the collection do not play any role in those relationships whatsoever because no such information is included
in the computation.

\subsection{Colonization and Language: Portugal, Spain and South America.}  

\begin{figure}[h!]
\includegraphics[width=0.55\textwidth]{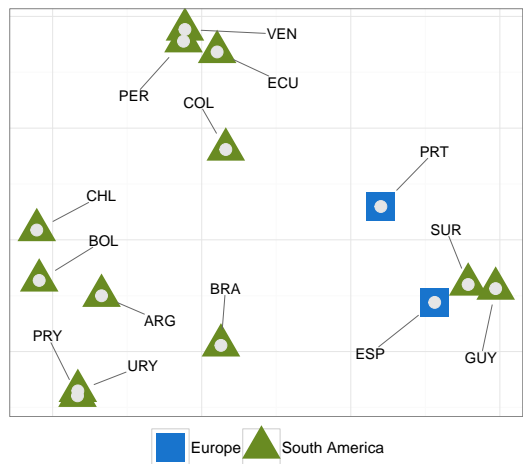}
\centering
\captionof{figure}{The map applied to Spain, Portugal and South America.}
\label{language}
\end{figure}

This example is motivated by colonization history
and deals with the subgraph consisting of Spain, Portugal and all of South America. The outcome is a very interesting clustering: the two clear outliers are Guyana (colonized by
the Dutch but then came under British rule and is now a member of the Commonwealth with English as official language) and Suriname (colonized
by the Dutch and using Dutch as an official language). The map clearly reveals those two countries to behave as if they were European with respect to their trading behavior
with Portugal and Spain and the rest of South America. We see that Brazil (the sole American country with Portuguese as official language) is somewhat at a distance from the remaining Spanish-speaking 
countries but not as isolated as one might think -- it seems reasonable that given its size and central location, these geopolitical factors are more important
than linguistic-historical factors.

\section{Conclusion}
We have presented a mathematical tool that maps a collection of countries into Euclidean space $\mathbb{R}^2$ using only the amount
of mutual trade between these countries. We have shown that complex, multinational trade is indeed governed by relatively few different
factors (a small number compared to the total number of countries) and that these
low-dimensional factors encoded in the eigenvectors capture precisely the notion of ``distance'' that has been actively investigated in the context of the gravity equation.
Examples show that it has at least geographical, historical and linguistic components interacting in nonlinear and meaningful ways. We believe that our
approach can be helpful in providing an alternative approach to ``distance'' that works directly on the given data and does not require an external axiomatic
approach to the concept. 
The method clearly has a series of applications in econometrics as it allows to quantify complex multinational relationships as well as their evolution
and this we plan to address in future work.

 \subsection*{Acknowledgement.} S.S. and T. W. were partially supported by \#INO15-00038 from the Institute of New Economic Thinking (INET).
S.S. was supported by an AMS-Simons Travel Grant and a Yale Provost Travel Grant. The authors are grateful for valuable discussions with Xiuyuan Cheng, Alexander Cloninger, A. Stephen Morse and Peter K. Schott.

{\scriptsize
\bibliography{gravity,distance,network}}

\bibliographystyle{elsarticle-harv}

\end{document}